\documentclass{ws-procs975x65}

\usepackage{amssymb}
\usepackage{amsmath}
\usepackage{amsfonts}

\renewcommand\ln{\ell{n}}
\newcommand\beq{\begin{equation}}
\newcommand\eeq{\end{equation}}
\newcommand\bea{\begin{eqnarray}}
\newcommand\eea{\end{eqnarray}}
\newcommand\bseq{\begin{subequations}} %solo con amsmath
\newcommand\eseq{\end{subequations}}
\newcommand\bal{\begin{align}}  %% controllare se funzionano
\newcommand\ealign{\end{align}}    %% queste abbreviazioni per {align}
\renewcommand\ln{{\rm ln}}

\begin{document}

\title{Covariant Lyapunov Exponents for the Mixmaster}
			
\author{GIOVANNI IMPONENTE}

\address{Dipartimento di Fisica, 
 						Universit\'a ``Federico II'', 
 						Napoli  and %\\
					 INFN -- Napoli -- Italy \\
					ICRA -- International Center for
								Relativistic Astrophysics \\
E-mail: imponente@icra.it}

\author{GIOVANNI MONTANI}

\address{Dipartimento di Fisica -- G9
					Universit\'a ``La Sapienza'', Roma -- Italy \\
 ICRA -- International Center for
								Relativistic Astrophysics \\
				E-mail: montani@icra.it}  


\maketitle

\abstracts{
The dynamics of the Mixmaster 
Universe is analized in a covariant 
picture via Misner--Chitre-like 
variables for an ADM  Hamiltonian approach. 
The system outcomes as isomorphic 
to a billiard on the Lobachevsky plane
and Lyapunov exponents are calculated 
explicitly. 
}

\section{Introduction}

One of the most outstanding problem discussed 
during the 1990s regarding the Mixmaster 
 \cite{BKL70,M69} cosmology concerned the 
 covariant nature of its chaoticity \cite{H94}.
The subtleties arose in this debate because of 
the non-covariant nature of the standard chaos 
indicators when applied to General Relativity
\cite{B93,BBT,FFM91}.
The peculiarity of the Mixmaster model is due to 
the vanishing of its Hamiltonian and to the 
non-positive character of the corresponding 
kinetic term. 
Reliable indications about the chaos covariance 
 appeared in later literature 
\cite{CL97,ML00,IM01}. 

Here we show that the Mixmaster
is isomorphic to a 
billiard ball on a Lobachevsky plane and that this 
picture is independent on the lapse-function choice;
therefore we can apply the standard Lyapunov exponents 
to characterize the chaos covariance.
We base our analysis over the existence of an 
asymptotic energy-like constant of motion which 
permits the definition of a standard Jacobi metric 
for the Arnowitt--Deser--Misner (ADM)
reduced model \cite{IM01}.

\section{Covariant ADM Reduction}

Starting from Misner variables 
$\alpha, \beta_+, \beta_-$ \cite{MTW}
we define the new Misner--Chitre-like ones
$f(\tau),\xi,\theta$
\begin{align}
\label{f2}
%&
\alpha = -e^{f\left(\tau\right)}\xi \,, \qquad% \\
\beta_+ = e^{f\left(\tau\right)}\sqrt{\xi^2 -1}
\cos \theta \,, \qquad %\\
\beta_- = e^{f\left(\tau\right)}
\sqrt{\xi^2 -1}\sin \theta  \,, 
\end{align} 
with $f$ denoting a generic functional form of $\tau$, 
$1\le \xi <\infty$ and $0\le \theta < 2 \pi$. \\
In terms of (\ref{f2}) the ADM reduction
of the Mixmaster dynamics provides the action
\begin{equation} 
 \mathcal{S}_{\textrm{RED}}=\int 
 \left( p_{\xi}d \xi  +  p_{\theta} {d\theta}
 -  
 \sqrt{\varepsilon^2+U} ~ {df} \right)\, ,
\label{q} 
\end{equation} 
where $p_{\xi}$ and $p_{\theta}$ denote the conjugate 
momenta to $\xi$ and $\theta$, respectively, 
\beq
\varepsilon ^2 = \left({\xi}^2 -1\right){p_{\xi}}^2 
+\frac{{p_{\theta}}^2}{{\xi}^2 -1} 
\label{d2}
\eeq
and $U(\xi,\theta,\tau)$ can be modelled near
the singularity $(\tau \rightarrow \infty)$ 
by the infinite walls 
\beq
U= \sum_i \Theta_{\infty}(H_i) \, , \quad 
	\Theta_{\infty}(x)= \left\{ \begin{array}{l}
												0 			\quad x>0 \\
												\infty  \quad x\leq 0\, ;
												\end{array}
												\right. 
\label{pinf}
\eeq
the \textit{anisotropy para\-me\-ters} 
$H_i$ $(\sum_i H_i =1)$ in the 
Misner--Chitre-like variables, read 
%\bseq
\bal
%& 
H_{1,2} = \frac{1}{3} - \frac{\sqrt{\xi ^2 - 1}}{3
		\xi }\left(\cos\theta \pm
		 \sqrt{3}\sin\theta \right)  \, , \qquad 
%& H_2 = \frac{1}{3} - \frac{\sqrt{\xi ^2 - 1}}{3
%		\xi }\left(\cos\theta - \sqrt{3}\sin\theta \right)  \\ 
 H_3 =  \frac{1}{3} + 2\frac{\sqrt{\xi ^2 - 1}}{3\xi }
       \cos\theta \, ,
\label{hs}
\end{align}
%\eseq
which do not depend on the (time) variable $\tau$. \\
The domain $\Gamma_H$ where $U$ vanishes is dynamically
closed and, within it, $\varepsilon$ behaves as a constant
of motion, i.e. 
$	\frac{d\varepsilon}{d\tau}=
\frac{\partial\varepsilon}{\partial\tau}=0$, which 
implies $\varepsilon=E=\textrm{const.}$. \\
The covariance of this picture is ensured 
because of the time-gauge relation
\begin{equation} 
N\left(\tau\right)= \frac{12 D}{E} 
e^{2f}  \frac{df}{d\tau}  \, ,
\label{rs} 
\end{equation} 
in which 
$D= \exp( -3 \xi e^{f\left(\tau\right)} )$.
The fixing of  a specific time variable 
corresponds to choose a suitable function 
$f (\tau)$.

\section{Lyapunov Exponents}

In $\Gamma_H$ the dynamics of the Mixmaster is 
summarized by the variational principle
\beq
\delta \int \left(p_\xi d\xi +
	p_\theta d\theta\right) =0 \, ;
\eeq
this picture can be restated in terms of 
a billiard  geodesic flow on the Lobachevsky 
plane described by 
the line-element \cite{IM01}
\begin{equation}
dl^2 =E^2 \left[ 	\frac{d{\xi }^2}{{\xi}^2 -1}+  
\left(\xi^2 -1\right) d {\theta }^2 \right] \, .
\label{mm}
\end{equation}
Such a space has a constant negative curvature
(the Ricci scalar is equal to $-2/E^2$) and 
the dynamical instability 
is studied via the geodesic deviation equation
as projected over the Fermi basis
\bseq
\begin{align} 
v^i &=\left(\frac{1}{E}\sqrt{{\xi}^2-1}\cos{\phi},\, 
		\frac{1}{E}\frac{\sin{\phi}}{\sqrt{{\xi}^2-1}}\right)  \\
w^i &=\left(-\frac{1}{E}\sqrt{{\xi}^2-1}\sin{\phi},\,
		\frac{1}{E}\frac{\cos{\phi}}{\sqrt{{\xi}^2-1}} \right) \, ,
\label{nn}
\end{align} 
\eseq
where $\phi(\tau)$ lies in the interval
$[0, 2\pi)$;
the vector $v^i$ is the geodesic field
while $w^i$ is parallely transported 
along it.\\
Projecting the geodesic deviation equation along 
the vector $w^i$ (its component along the geodesic 
field $v^i$ does not provide any physical information 
about the system instability), the corresponding 
connecting vector (tetradic) component $Z$ satisfies 
the equivalent equation 
\begin{equation}
\frac{d^2 Z}{ds^2}=\frac{Z}{E^2} \, .
\label{qqz}
\end{equation}
Expression (\ref{qqz}),
as a projection on the tetradic 
basis,  is a scalar and therefore completely 
independent of the choice of the variables.
Its general solution reads
\begin{equation}
Z\left(s\right)=c_1 e^{\frac{s}{E}}+
c_2 e^{-\frac{s}{E}} \, , \qquad 
		c_{1,2}=\textrm{const.} \,  ,
\label{rr}
\end{equation}
and the invariant Lyapunov exponent 
defined by 
\begin{equation}
\lambda_v =\sup \lim_{s\rightarrow \infty} 
\frac{\ln\left(Z^2+ \left(\frac{dZ}{ds} 
								\right)^2\right)}{2s} \, ,
\label{ssz}
\end{equation}
takes the value 
\begin{equation}
\lambda_v =\frac{1}{E} > 0 \, .
\label{tt}
\end{equation}
When the point-universe bounces against the 
potential walls, it is reflected from a geodesic 
to another one thus making each of them unstable. 
Though up to the limit of our potential wall 
approximation, this result shows without any 
ambiguity that, independently of the choice of 
the temporal gauge, the Mixmaster dynamics is 
isomorphic to a well described chaotic system.

\end{document}